\documentclass[11pt,twoside]{article}
\usepackage{./asp2014}
\usepackage{graphicx}
\usepackage{natbib}

\resetcounters

\begin{document}

\title{Molecular Clouds in Galaxies}
\author{Adam K. Leroy,$^1$ Alberto Bolatto,$^2$ Erik Rosolowsky,$^3$ and Eva Schinnerer$^4$}
\affil{$^1$Ohio State University, Department of Astronomy, Columbus, OH 43210; \email{leroy.42@osu.edu}}
\affil{$^2$ Department of Astronomy, University of Maryland, College Park, MD 20742}
\affil{$^3$ University of Alberta, Department of Physics, 4-183 CCIS, Edmonton AB T6G 2E1, Canada}
\affil{$^4$ Max Planck Institute f\"ur Astronomie, K\"onigstuhl 17, 69117, Heidelberg, Germany}
\paperauthor{Adam Leroy}{leroy.42@osu.edu}{}{Ohio State University}{Department of Astronomy}{Columbus}{OH}{43210}{USA}

\begin{abstract}
Stars form in cold clouds of predominantly molecular (H$_2$) gas. We are just beginning to understand how the formation, properties, and destruction of these clouds varies across the universe. In this chapter, we describe how the thermal line imaging capabilities of the proposed next generation Very Large Array (ngVLA) could make major contributions to this field. Looking at CO emission, the proposed ngVLA would be able to quickly survey the bulk properties of molecular clouds across the whole nearby galaxy population. This includes many unique very nearby northern targets (e.g., Andromeda) inaccessible to ALMA. Such surveys offer a main observational constraint on the formation, destruction, lifetime, and star formation properties of clouds. Targeting specific regions, the ngVLA will also be able to heavily resolve clouds in the nearest galaxies. This will allow detailed studies of the substructure and kinematics --- and so the internal physics --- of clouds across different chemical and dynamical environments.
\end{abstract}

\section{The Fundamental Unit of the Star-Forming Interstellar Medium}

Stars form in massive, cold clouds of predominantly molecular (H$_2$) gas \citep[e.g., see reviews by][]{FUKUI10,KENNICUTT12,HEYER15}. Understanding the formation, evolution, and destruction of these giant molecular clouds (GMCs) is a major focus of millimeter wave astronomy, and we are still learning how their properties are imprinted on the resulting stellar population.

In theory, these clouds form through a mixture of large scale instabilities, conversion of atomic to molecular gas, and agglomeration of smaller structures (e.g., see Dobbs et al. 2014). Their internal motions are dominated by supersonic turbulence, and they have a large amount of sub-structure. For example, local clouds have mean H$_2$ column densities of $n_{\rm H2} \sim 10^2$~cm$^{-2}$ but stars form in dense filaments and cores with $n_{\rm H2} > 10^5$~cm$^{-2}$ \citep[e.g., see][]{LADA10,ANDRE14}. In our current understanding, over one or more crossing or collapse times, a mixture of turbulence and external triggers collect low density gas into high density sub-structures. Then stars form inside these dense sub-structures. Stellar feedback from these newly formed stars then deposits energy and momentum back into the parent cloud, perhaps even destroying it.

But these broad statements cover a large amount of ignorance. Many of our basic questions about GMCs still lack satisfactory answers. These include:

\begin{enumerate}
    \item What is the dominant formation mechanism for GMCs in different environments?
    \item How much molecular gas is actually in bound structures, and how much exists in a ``diffuse'' non star-forming phase?
    \item How do the dynamical state, character of turbulence (e.g., strength, driving scale, and type of driving), internal structure, and chemical state of GMCs depend on environment? How do they evolve with time?
    \item Are the star-forming sub-structures within GMCs created mainly by turbulence or by interactions with the large-scale environment (e.g., triggered by cloud-cloud collisions or spiral arm passage)?
    \item What is the main destruction mechanism for GMCs? Stellar feedback, galactic gravitational forces, or something else? How long do they live once they have been formed?
    \item Does the fraction of gas that a GMC turns into stars over its lifetime depend on its properties? How does the rate at which stars form depend on the clouds' properties? What about the fraction of stars formed in bound star clusters?
\end{enumerate}

The key to progress in these areas are observations that measure the demographics and internal structure of GMCs over the full diversity of dynamical, energetic, and chemical environments found in the universe. We aim to understand how GMCs work in environments spanning from low metallicity, shallow potential well dwarf galaxies to violent, merger-induced starbursts or the highly stable gas disks in early-type and post-starburst galaxies. 

In each of these environments, complete demographics of the cloud population will allow astronomers to constraint the life cycle, destruction, and formation mechanisms of clouds. Meanwhile, detailed chemical, structural, and kinematic observations of individual clouds offer the prospect to to illuminate the physics at play inside these crucial objects.

\begin{figure}[t!]
\begin{center}
\includegraphics[width=0.45\textwidth]{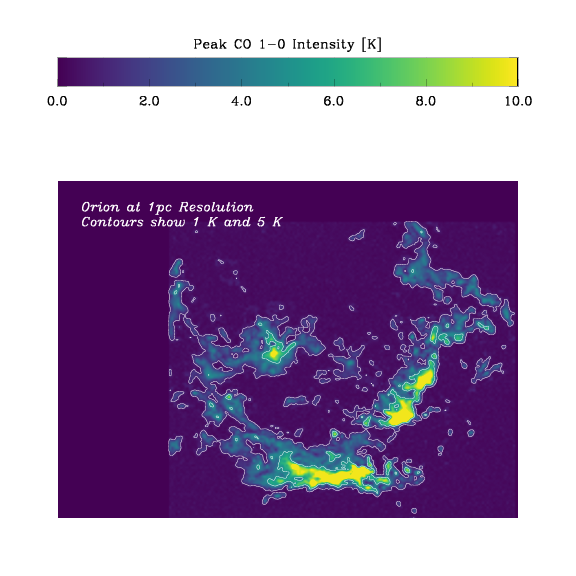}
\includegraphics[width=0.45\textwidth]{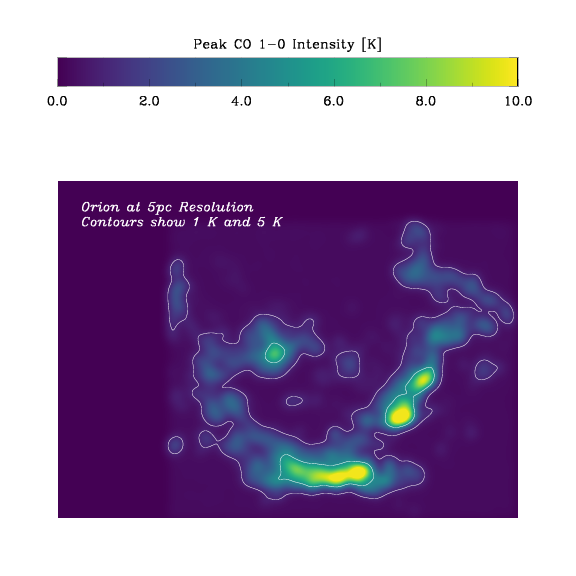} \\
\includegraphics[width=0.45\textwidth]{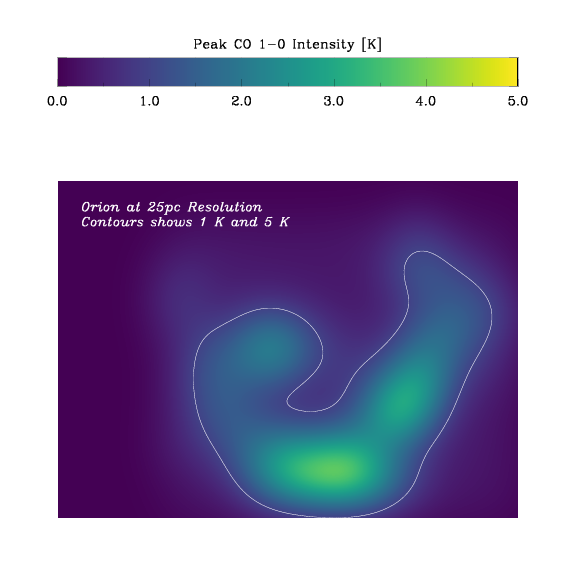}
\includegraphics[width=0.45\textwidth]{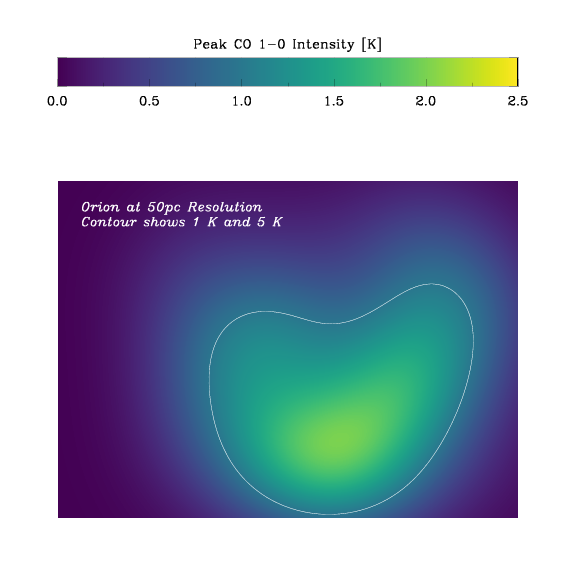}
\end{center}
\caption{{\bf The ngVLA View of Clouds in Distant Galaxies --- Orion at 1, 5, 25, and 50~pc Resolution:} The Orion Molecular clouds, as seen in CO $J=1\rightarrow0$ by \citet{DAME01}. ({\em Top Left:}) At 1 pc resolution, as the ngVLA could see molecular clouds in Local Group galaxies. The contours show $1$ and $5$~Kelvin brightness temperature. The ngVLA will achieve $\sim 1$~K rms noise per 10~km~s$^{-1}$ channel at $0\farcs1$ resolution in $1$ hour on source, so that this detailed view of clouds (in fact $\sim 2$ to $3$ times sharper than this) will be possible for Local Group targets, including the spiral galaxies M33 and M31. ({\em Top Right:}) A similar view at $5$~pc resolution, resolution readily achievable in the nearest groups of galaxies, e.g., for the M81+M82 group. These high resolutions will allow detailed study of how cloud subtructure and internal kinematics change as a function of environment. ({\em Bottom Left} and {\em Bottom Right:}) Progressively lower resolution views. The bottom two panels, at resolution 25~pc and 50~pc show the view the ngVLA can achieve in {\em $\sim 1$ minute integration time} targeting galaxies at $\sim 5{-}20$~Mpc. This allows the fantastic possibility of cloud demographics across the whole local galaxy population. This figure the peak brightness temperature structure, but spectroscopic imaging also returns information on the cloud kinematics, giving access to bulk energetics (at low resolution) and diagnostics of turbulence and internal flows (at high resolution).}
\end{figure}

\section{Millimeter-Wave Spectroscopic Imaging as the Key to Understand GMCs}

High physical resolution millimeter spectroscopic imaging is the key to understand GMCs. Specifically, the low-$J$ rotational lines of CO, the second most abundant molecule (after the effectively invisible H$_2$), trace the mass, structure, and kinematics of GMCs. CO imaging has provided most of the basis for what we currently know about GMCs in other galaxies, and this situation seems likely to hold heading in to the next decades.

The problem is that isolating GMCs requires resolution of $\approx 50{-}100$~pc, while resolving clouds and measuring their internal structure requires resolution $\approx 1{-}10$~pc. CO line emission from a GMC has brightness temperature of order $\sim 1$~K and line width $\approx 1{-}20$~km~s$^{-1}$ \citep[e.g., see][]{SUN18}. At the $\approx 5{-}20$~Mpc distance of the nearest massive, star-forming galaxies, this translates to a required resolution of $\sim 0\farcs5{-}2\arcsec$ to isolate GMCs and study their population. And higher resolution of $\sim 0\farcs1{-}0\farcs4$ is required to resolve GMC substructure even in these nearby spiral galaxies. For rarer systems that lie at greater distances (e.g., major galaxy mergers and rare phases of galaxy evolution) the angular resolution requirements will be even stricter.

Interferometers suffer from a well-known tradeoff between surface brightness sensitivity and resolution, and for the pre-ALMA generation of interferometers, imaging a $\sim 1$~K line at these resolutions was daunting. Heroic efforts by the pre-NOEMA PdBI obtained the first cloud-scale ($\theta = 1'' = 40$~pc) map of CO of M51 \citep[e.g.,][]{SCHINNERER13} and CARMA imaged the inner regions of many nearby galaxies \citep{DONOVANMEYER13}. But these efforts required large time investments (dozens to hundreds of hours of the highest quality time available to the telescopes) and often remained sensitive only to the most massive GMCs. 

These pre-ALMA studies already conclusively demonstrate that the classic picture of a universal population of GMCs does not hold across all environments. Rather, GMCs appear to vary in their surface density, density, and line width \citep[e.g.,][]{HEYER09,HUGHES13B,DONOVANMEYER13}. And the data strongly suggest coupling between the local disk structure, ISM conditions, and the properties of clouds \citep[e.g.,][]{FIELD11,HUGHES13B}.

In the ALMA era, this picture has become even clearer. Observations show real, significant variations of GMC-scale properties across the local universe \citep[e.g., see][among many others]{LEROY15A,DAVIS17A,SUN18,EGUSA18}. The primary sense of this variation appears to be huge variations in the internal pressure of molecular gas \citep{SUN18}, but more modest -- though still crucial -- changes in the dynamical state of gas are also evident \citep{DAVIS17A,EGUSA18,SUN18}. In addition, the mass function of clouds has long been known to vary within and among galaxies \citep[e.g.,][]{ROSOLOWSKY05,COLOMBO14A,FREEMAN17}

These observations drive the need for large-scale comparative demographic studies of clouds. They also underline the need for detailed studies of clouds in different environments -- if the populations of clouds changes from place to place, do the governing physics also change? How do the turbulence and sub-structure compare across these different cloud populations? And how do the stars and clusters formed vary as the density, turbulence, and dynamical state of the ISM also vary?

\section{The Next Generation Very Large Array --- Molecular Cloud Populations Well Beyond Virgo}

\begin{figure}[t!]
\begin{center}
\includegraphics[width=0.49\textwidth]{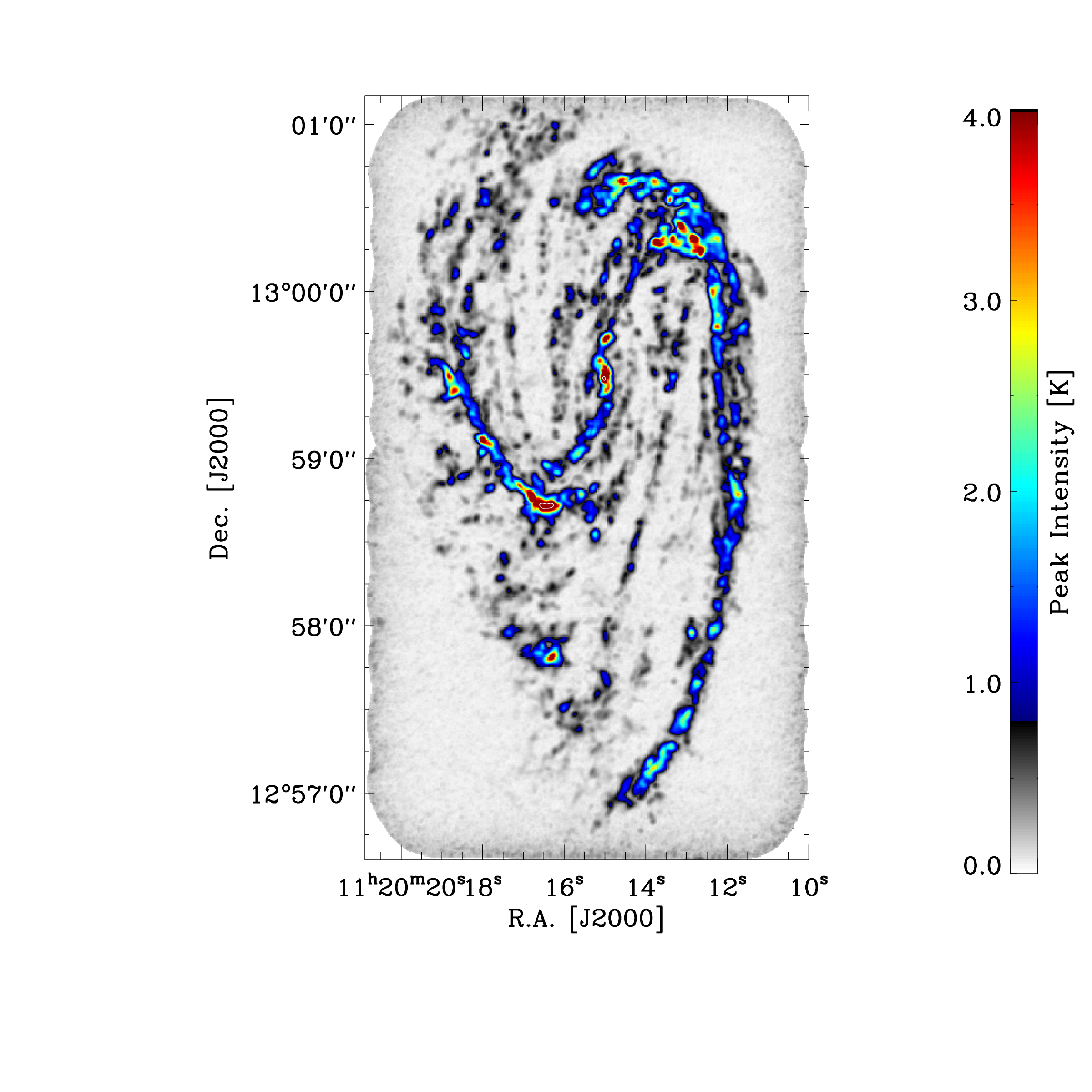}
\includegraphics[width=0.49\textwidth]{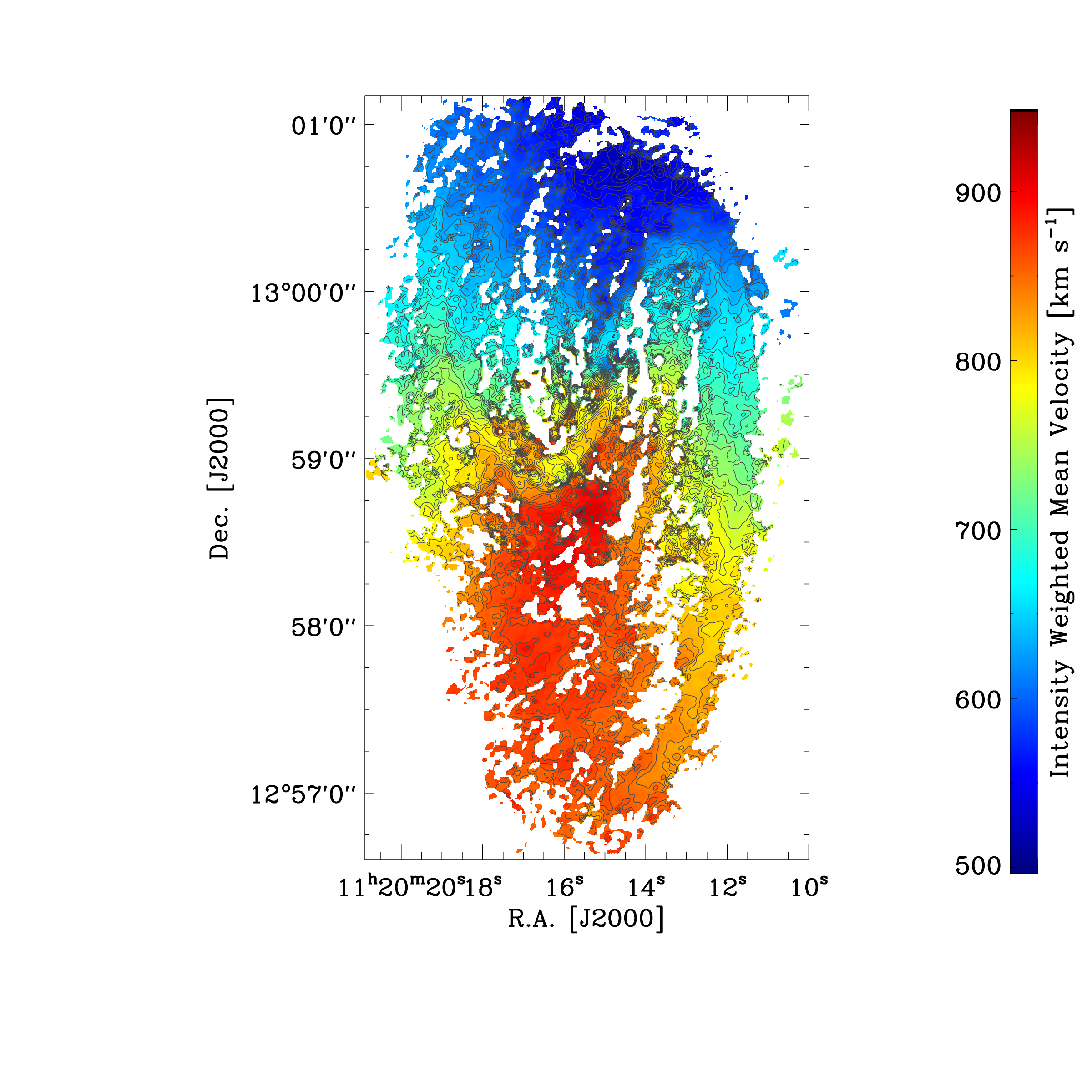}
\end{center}
\caption{{\bf The ngVLA View of Clouds in Distant Galaxies --- Whole Galaxies at $\sim 50$~pc Resolution.} Peak CO~(2-1) intensity ({\em left}) and intensity-weighted velocity ({\em right}) emission from the nearby barred spiral NGC~3627 (Messier 66) at 50~pc ($\approx 1.3''$) resolution. At this resolution, the gas in the galaxy breaks up into individual cloud-sized patches of emission and the line widths, masses, energetics of individual clouds can be estimated. \citet{SUN18} used these data to show a strong dependence of line width, surface density, and internal pressure on dynamical environment across the galaxy and variations among galaxies. Targeting the fundamental CO~(1-0) transition, the ngVLA can map a field to comparable sensitivity in less than a minute. This allows surveys of clouds across a whole big galaxy in only a few hours out to beyond the Virgo Cluster, and farther using the ngVLA in high resolution modes. This ability is currently new with ALMA, and is ushering molecular cloud studies into the era of survey science. {\em The ngVLA will build on ALMA's progress in this field, observing the fundamental transition, targeting unique nearby northern targets, and mapping with potentially a few times faster speed than ALMA}.}
\end{figure}

Answering these questions requires observing CO, or similar, line emission with sub-Kelvin sensitivity at angular resolutions $\theta = 0\farcs1{-}1\arcsec$. This requires a massive, sensitive, extended-configuration interferometer like the proposed next-generation Very Large Array \citep[ngVLA;][]{CARILLI15}. Because CO emission has significant power at spatial scales reaching up to many hundreds of pc \citep[e.g.,][]{PETY13}, these observations require good sensitivity at all spatial scales, including short spacing information.

The thermal line imaging capabilities of the ngVLA are exactly suited to this kind of science. In a single field in an hour, the ngVLA reaches rms noise of 0.01~K at $\theta=1''$ over a 10~km~s$^{-1}$ line width, typical for a GMC\footnote{Taking the 93~GHz values for the reference design to apply to CO~$J=1\rightarrow0$ at 115~GHz.}. This allows the prospect to reach $\sim 0.1$~K sensitivity in under a minute. This will allow the ngVLA to easily mosaic many hundreds of fields in a few hours, measuring the CO surface brightness, luminosity, and line width of all clouds across a galaxy.

Capable of a noise of $\sim1$\,K in one hour at $\theta\approx0\farcs1$, the ngVLA can resolve the molecular cloud populations in galaxies at distances of $20-100$~Mpc. This will allow us to study the GMCs in environments vastly different from the ones present in our own Galaxy, such as Luminous and Ultra-Luminous Infrared Galaxies and galaxies in the nearest galaxy clusters, including those undergoing quenching and gas loss processes such as ram pressure stripping. 

This allows the prospect to statistically characterize how GMC populations depend on environment. Does the mass function of clouds shift in response to the local Jean's mass or the mean pressure of the ISM? Does the dynamical state (ratio of kinetic to potential energy) change between galaxy centers, regions of strong gas flow like bars or arms \citep[e.g.,][]{MEIDT13,EGUSA18}, and more quiescent disk galaxies? How do the mean density and mean internal pressure of clouds vary in response to changing disk conditions?

All of these questions can be effectively addressed with high completeness (and thus high sensitivity) surveys that estimate the mass, kinetic energy, and size of a patch of the molecular interstellar medium on the scale of an individual GMC. To be effective, these surveys need to cover large, well-characterized parts (or all area) of many galaxies, so that any individual field must be covered quickly. For this application, reaching resolutions approximately matched to GMCs is more important than heavily resolving them, and here the ngVLA ``core'' will be incredibly powerful. There are a few hundred moderately massive northern star forming galaxies within $\sim 20$~Mpc. At a few hours apiece, the ngVLA could survey CO emission at cloud scale from all of them. Expanding to cover all of the local (but more distant) starbursts \citep[e.g., the GOALS targets;][]{ARMUS09} and a sample of nearer, low metallicity dwarf galaxies \citep[e.g., many of the LVL targets;][]{LEE09} would add another few hundred hours. With order $\sim 1,000$ hours, the ngVLA would build definitive atlas of GMCs across all local environments.

Beyond just empirically understanding the influence of environment on clouds, these population studies are the crucial element to constrain the physics governing the formation, destruction, and evolution of clouds. In the last few years there has been large interest in resolving the ISM down to the scales of individual star-forming regions and using tracers that probe different phases of the star formation and feedback process \citep[e.g.,][]{KAWAMURA09,SCHRUBA10,KRUIJSSEN14,CORBELLI17}. With the ngVLA it will, for example, be possible to image a whole galaxy at cloud-scales using CO to trace molecular gas, HCN and HCO$^+$ to trace dense gas, 21-cm H{\sc i} line emission to trace atomic gas, and free-free continuum emission to trace young stars and ionized gas. The joint distributions of these tracers at high resolution are the key observation that encodes the life cycle of GMCs: their birth, life (including the formation of stars), and ultimate destruction.

Because of the wide instantaneous bandwidth of the ngVLA, such a survey would also be able to trace the chemical and physical conditions inside each cloud. Although the exact receiver set-up is not decided upon yet, simultaneous coverage of CO isotopologue lines ($^{13}$CO, C$^{18}$O, or even C$^{17}$O) could offer constraints on bulk gas conditions, optical depth, and maybe even nucleo-synthesis \citep[e.g.,][]{SLIWA17}. Covering high critical density lines (see gas density-focused white papers in this book) would provide probes of how gas density depends on cloud properties (or on time, as above). And comparing, both statistically and directly, the properties of clouds to local tracers of recent star formation will allow a direct link between gas properties, star formation efficiencies, star formation rates, and star formation timescales.

\begin{figure}[t!]
\begin{center}
\includegraphics[width=0.75\textwidth]{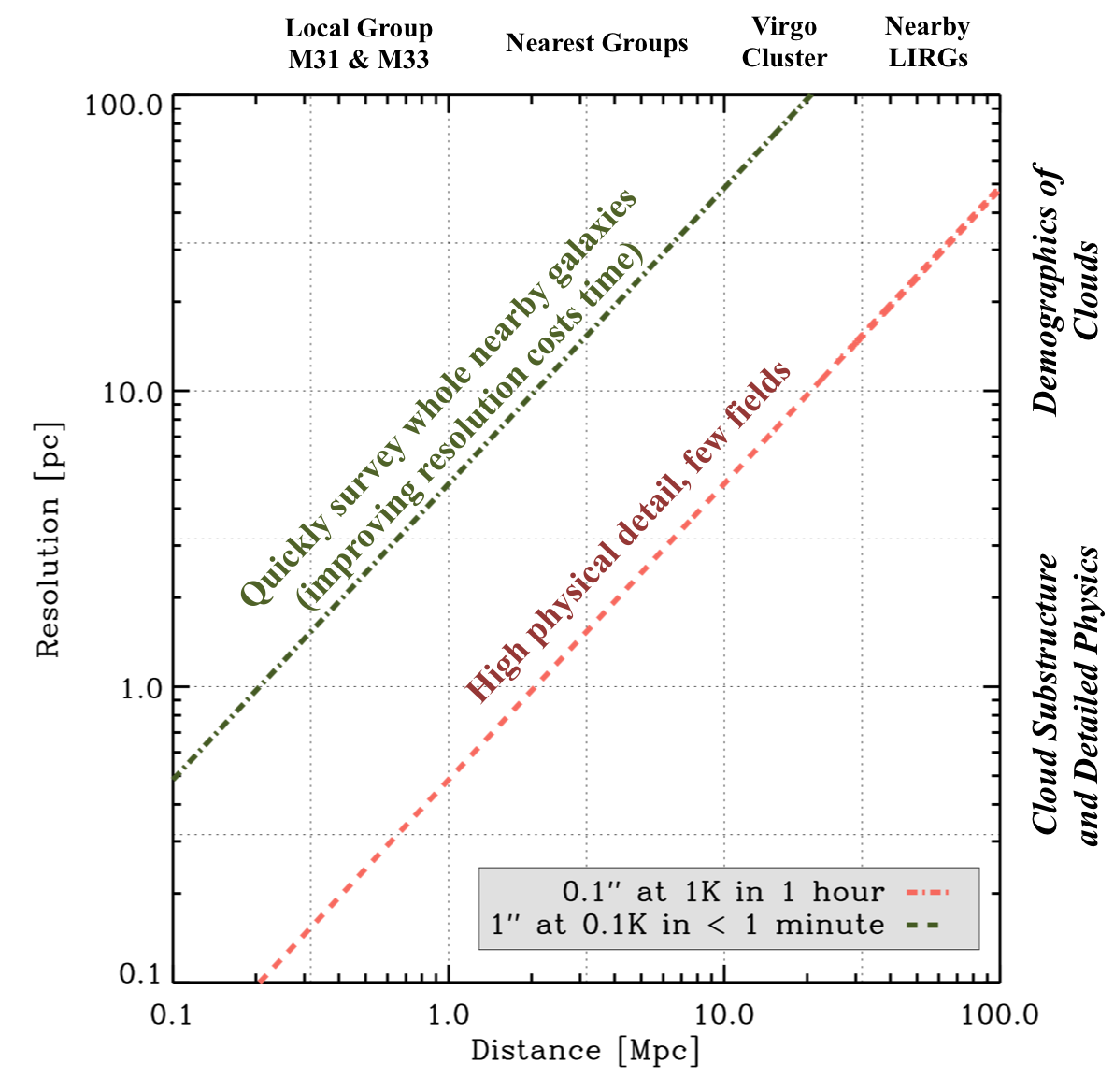}
\end{center}
\caption{{\bf Schematic View of ngVLA Capabilities Targeting Molecular Clouds.}. Physical resolution ($y$-axis) achieved at $1''$ (green) and $0\farcs1$ (red) resolution according to the ngVLA reference design. The green line shows a mode in which the ngVLA can survey whole large areas of nearby galaxies in a few areas. This will allow detailed studies in the Local Group and cloud demographics out to beyond the Virgo Cluster. The red line shows higher resolution mapping, still leveraging the ngVLA core, that can be used to study a few fields at high detail in a few hours. This will be crucial to study the internal physics of clouds in exquisite detail in the nearest galaxies. It is also the key mode to survey bright, but distant targets like LIRGs, which are rare systems that probe crucial extreme physics.}
\end{figure}

\section{The Next Generation Very Large Array --- The Local Group in Stunning Detail}

When focusing on a single field, the ngVLA reaches rms noise $\sim 1$~K at $\theta=0\farcs1$ resolution in 1 hour. This provides the prospect to dissect individual clouds in stunning detail out to many Mpc -- complementing the survey science described above with detailed studies of the physics playing out inside individual clouds. This also allows the Local Group spirals M31 and M33 to be observed with detail currently reserved for Milky Way molecular clouds \citep[similar to what is currently done by ALMA targeting the Magellanic Clouds, e.g., see][]{WONG17}. Conversely, large surveys of Local Group galaxies at $\theta \approx 1''$ would yield $\approx 4$~pc linear resolution over huge areas.

Heavily resolving the internal structure of molecular clouds in other galaxies is \textit{terra incognita} and offers a huge opportunity to gain insight into how the physics of star formation change (and stay the same) across the universe. Resolution $\theta = 0\farcs1$ translates to $\sim 0.5$~pc at $1$~Mpc. The ngVLA would be able to measure the internal (turbulent) size-linewidth relation \citep[e.g.,][]{HEYER04} as well as more sophisticated turbulence diagnostics \citep[e.g.,][]{BURKHART15} in clouds across M31, M33, the Local Group dwarfs, and (with somewhat more effort) galaxies in the nearest groups (including M81 and M82, $d\sim 3.5$~Mpc). The same observations would help constrain the column density distribution function, which has been the topic of intense interest in the Milky Way in recent years \citep[e.g.,][]{KAINULAINEN09,LOMBARDI15} because of its close link to star formation. 

These hyper-resolved observations would offer the prospect to search other galaxies for dense filamentary structures that have captivated Milky Way astronomers both within clouds \citep[e.g.,][]{ANDRE14} and on larger scales \citep[][]{JACKSON10}. These filaments have been argued to be fundamental to star formation \citep{ANDRE14} and perhaps to serve as the ``bones'' of the Milky Way's interstellar medium \citep{ZUCKER17}. Other studies have emphasized biases in earlier analysis \citep{PANOPOULOU17}, questioning whether these structures are indeed universal. An external, hyper-resolution ($< 1$~pc) view of GMCs in other galaxies offers huge potential to tell us how the cold, dense ISM is structured.

\section{Relationship to Other Facilities --- Building on ALMA's Legacy and a Close Complement to {\it JWST}}

Studying higher $J$ CO lines, CO~$J=2\rightarrow1$ and $J=3\rightarrow2$, ALMA is currently revolutionizing this field. Its capabilities in these higher $J$ transitions resemble those of the ngVLA in CO~$J=1\rightarrow0$. As discussed above, ALMA's findings so far suggest that there is significant depth to this topic, and in the nearby universe the north holds a set of key, very nearby targets. So the ngVLA is highly likely to pick up an active field and push it farther, exploring the effects of excitation and digging in to a new set of key targets. The Local Group spirals M33 and M31 stand out. These are the nearest star-forming spiral galaxies to the Milky Way. Because of this proximity, they will always be the place where one can explore ISM physics at high resolution and in greater detail than in more distant targets. Along similar lines, the M81 group, also too far north to be seen by ALMA, represents the nearest major violent interaction. It hosts the prototype merger-induced starburst M82. And M51, M101, and NGC 6946 have all been key targets to study gas and star formation due to their proximity and activity. But sit at declination $\delta >30^\circ$ (M33 may be surveyed by ALMA, the others are all at $\delta >40^\circ$).

In the north, the main other relevant facility will be NOEMA, which can and will pursue many of the same science goals. Establishing the basic structure and evolution of the molecular ISM is likely to be such a deep vein that all three telescopes (ALMA, NOEMA, and the ngVLA) can be expected to make major contributions.

In the early 2020s, NASA's {\em James Webb Space Telescope} ({\it JWST}) will launch. By observing near-IR recombination lines and the mid-IR dust continuum, {\it JWST} will be an unparalleled instrument to survey the earliest, most heavily embedded phases of star formation at high angular resolution. In this sense {\it JWST} and the ngVLA are extraordinarily complementary: ngVLA will observe the cold, star-forming gas and {\it JWST} will observe how it forms stars.

\section{Likely Impact}

With these capabilities, the ngVLA would immediately become the premier instrument to study molecular cloud populations and structure in other galaxies. It would represent a powerful northern complement to ALMA. Its observations would naturally complement observations of recent, embedded star formation by the {\em JWST} (as well as ngVLA continuum observations of recent star formation).

\acknowledgements AKL is partially supported by the National Science Foundation under Grants No. 1615105, 1615109, and 1653300.




\end{document}